\documentclass[12pt,a4paper]{article}
 \usepackage{amsmath,amstext,amssymb,amscd}
 \usepackage{graphicx}

\begin{document}

\title{A short way to stability}

\author{Danilov V.I. \thanks{Central Economics and Mathematics Institute of RAS, Moscow,Russia; e-mail: vdanilov43@mail.ru}, Shamkanov D.S. \thanks{Mathematical Institute of RAS, Moscow, Russia; e-mail: daniyar.shamkanov@gmail.com}}
\date{\today}
\maketitle

\begin{abstract}

The paper presents a concise and straightforward approach way to obtain the basic facts about stable contract systems in bilateral markets: existence, polarization, and lattice structure. The primary tool employed is the notion of desirability operator. This notion is essentially equivalent to the concept of Plott choice function, but it allows for a clearer visualization of stable systems of contracts as fixed points of some operator. The existence and other properties of these fixed points is obtained using Knaster-Tarski fixed-point theorem.

Additionally, we explore related topics such as order structure of the set of stable contract systems, the polarization of the interests of the parties, the constructiveness of the approach, its application to comparative statics, a relaxation of bipartition assumption, as well as generalization to systems where contracts can be established with varying intensities instead of being either accepted or denied.

\textbf{Keywords:} stable systems of contracts, graduated contracts, Blair's order, desirability operators, fixed points, lattices, Knaster-Tarski theorem.

\end{abstract}

The article's most accurate title should be: "A Concise Path to the Stability Theory in Bilateral Markets".  The paper almost does not introduce new facts but offers a streamlined approach to deriving key theory results: the existence of stable contract systems, the polarization of interests of the two sides of the market, and the lattice structure of the set of stable systems. This method remains effective even in the more general scenario of graduated contracts.

This succinctness and clarity are achieved through the concept of the desirability operator (Section 2). The preferences of market participants are defined using Plott choice functions; essential details about these choice functions are collected in Section 2. These choice functions are transformed into desirability operators.

The idea of stability of contract system is introduced in Section 1. Here is a brief overview: the participants involved in future contracts are divided into two distinct groups, with contracts only possible between members of opposing groups. These groups are conventionally referred to as firms and  workers, though they could represent any dual sets of participants, such as colleges and students, banks and depositors, or grooms and brides. Each firm $f$ and each worker $w$ has a set $X(f,w)$ of potential contracts they can enter into. In principle, a firm or worker can negotiate with multiple counterparts, so their  "preferences" are expressed through choice functions. If a firm $f$ has access to a set $A$ within $X(f)=\coprod_w X(f,w)$, it selects a subset $C_f(A) \subseteq  A$. Here, $X(f)$ is the set of all contracts potentially available to firm  $f$. The same applies to workers. Any subset of $S$ within the total contract set $X=\coprod_{f,w} X(f,w)$ constitutes  a contract system. A contract system is called stable if no participant wishes to  exclude any its contract from this system, and no pair $(f,w)$ desires to conclude a new contract outside the system. This concept is elaborated more formally in Section 2.

It is clear that all companies can be considered as a single entity, and all workers as a single individual. The decision-making processes of these aggregated entities are defined over the set $X$, as demonstrated by \cite{F, Dan-25}. Consequently, in the subsequent discussion, we will assume the existence of only two entities: the Firm and the Worker. Their preferences over the set of contracts $X$ are represented by the choice functions $F$ and $W$, respectively. Throughout the following sections, it is presumed that these choice functions adhere to the Plott condition. In the next Section, we will present the necessary definitions and properties of Plott choice functions; for a more comprehensive explanation, we refer to \cite{Dan-25}.

\section{Stability}

Consider the following task. There is a set $V$ of agents who can enter into binary contracts (i.e. contracts between two agents). The set of potentially possible contracts between  agents $v$ and $w$ is denoted by $X(v,w)$; the (disjoint) union of these sets denoted as $X$. In fact, we are talking about (multi)graph $(V, X)$ with the set of vertices $V$ and the set of edges $X$.

An arbitrary subset $S\subseteq  X$ is called the \emph{contract system}; $S(v)$ stands
 for the intersection of $S\cap X(v)$. Contracts are entered into voluntarily; no one can be forced into a contract, and no one can stop a contract if both parties agree. The specific contracts that are formed depend on the agents' preferences for different contracts. Therefore, we need to describe the preferences of agent  $v$ on the set $X(v)=\cup_{w}X(v,w)$.

Since agents can enter into several contracts, they must compare subsets of contracts rather than individual ones. This naturally leads us to consider choice functions. A choice function
on a set $X$ is a mapping of $C:2^X\to 2^X$ such that $C(A)\subseteq  A$ for any $A\subseteq  X$. Given the set $A$, the agent selects a subset  $C(A)$ of "best" contracts for it.

Thus, the complete description of the problem is consists in indication, for each agent $v$,
a choice function $C_v$ on the set $X(v)=\cup_{w}X(v,w)$. A \emph{solution} to the problem is the so-called \emph{stable}  system of contracts, that is, a system of contracts $S$, which satisfies two requirements:

a. $C_v(S(v))=S(v)$ for any $v\in V$;

b. If the contract $x$ between agents $v$ and $w$ belongs to $C_v(x\cup S(v))\cap C_w(x\cup S(w))$,
then $x\in S$.

These two requirements express the voluntary nature of the conclusion of contracts. The first condition is called individual rationality - no one wants to abandon the proposed system of contracts $S$. The second condition says that no pair of agents wants to conclude a new contract that is desirable for them; this condition is called non-blockability.

The existence of stable contractual systems hinges on agents preferences and topology of the contract graph. Even in a simple scenario with just two agents, no solution may exist. Therefore, we will focus on cases where  choice functions of all agents satisfy the rationality condition proposed by Plott \cite{P}.

\section{Reminders of Plott choice functions}

A choice function $C$ on set $X$ is called an \emph{Plott function} if it
satisfies two conditions:

\emph{Consistency}. If $C(A)\subseteq  B\subseteq  A$, then $C(A)=C(B)$.

\emph{Substitutability}. If $A\subseteq  B$, then $A\cap C(B)\subseteq  C(A)$.

The first condition expresses the rationality of the agent's behavior. The second condition posits  that elements within the set $X$ are considered as substitutes for each other in an economic sense. If one item from a menu $A$  becomes unavailable, all the other selected items remain part of the choice. There are situations where the condition of substitutability is not fulfilled.

The substitution condition implies (and is actually equivalent to) the following property: for any
$A,B\subseteq  X$

\begin{equation}\label{subad}
C(A\cup B)\subseteq  C(A)\cup B.
\end{equation}
\noindent Indeed, since $A\subseteq  A\cup B$, then from the substitutability $C(A\cup B)\cap A\subseteq  C(A)$. Therefore, $C(A\cup B)=C(A\cup B)\cap (A\cup B)=(C(A\cup B)\cap A) \cup (C(A\cup B)\cap B) \subseteq  C (A)\cup B$.
Applying the consistency condition to the inclusions
$$
C(A\cup B)\subseteq  C(A)\cup B\subseteq  A\cup B,
$$
we derive the following equality, first introduced by Plott in \cite{P} under the term "path independence"
\begin{equation}\label{Plott}
C(A\cup B)=C(C(A)\cup B).
\end{equation}
We refer to \cite{Dan-25} for examples of Plott functions,  which illustrate how this concept is a fascinating extension of partial order.\medskip

Now we introduce two derived concepts: the Blair (hyper)relation $\preceq =\preceq _C$ and the desirability operator $D=D_C$. They will be crucial in our later discussions.\medskip

\textbf{Definition.} \emph{Blair relation} $\preceq $ is given on the set $2^X$ and is defined as follows: for
subsets $A$ and $B$
$$
A\preceq  B\Leftrightarrow C(A\cup B)\subseteq  B.
$$
(Instead of $\subseteq  B$, one can write $=C(B)$.)

Obviously, $A\subseteq B$ implies $A\preceq B$. Another example: $A\preceq  C(A)$. Indeed,
according to (\ref{subad}), we have $C(A\cup C(A))\subseteq  C(A)\cup C(A)=C(A)$.\medskip

The relation $\preceq $ on $2^X$ is reflexive and transitive. It is sometimes called a relation of
\emph{revealed preference} for the following reason: the set of $A$ is available for selection from $A\cup B$, but only elements from $B$ are selected. The corresponding equivalence relation is denoted as $\approx$ or $\approx_C$. For example,
\begin{equation}\label{approx}
A\approx C(A),
\end{equation}
\begin{equation}\label{app} A\preceq  B \Leftrightarrow   A\cup B\approx B.
\end{equation}

\textbf{Definition.} An element $x\in X$ is called \emph{desirable in the state} $A\subseteq  X$,
if $x\in C(A\cup x)$. Typically, we write $A\cup x$ instead of $A\cup\{x\}$. The set of desired elements is denoted by $\widehat{C} (A)$. It is obvious that
\begin{equation}\label{ch}
C(A)=A\cap \widehat{C}(A).
\end{equation}
This equation reveals that the operator $\widehat{C}$ defines the choice function $C$.

Desirability operators have two important properties. The first property is:
\begin{equation}\label{inv}
\widehat{C}(A)=\widehat{C}(C(A))=\widehat{C}(A\cap \widehat{C}(A)).
\end{equation}
It follows from (\ref{Plott}) because $C(A\cup x)=C(C(A)\cup x)$.

The second property of the operator $\widehat{C}$ is its \emph{antitonicity}: $A\subseteq  B$ entails $\widehat {C}(B)\subseteq  \widehat{C}(A)$. It follows from the substitutability condition of choice function $C$. In fact, an even stronger statement is true:

\begin{equation}\label{anti}
\text{if } A\preceq  B \text{, then } \widehat{C}(B)\subseteq  \widehat{C}(A).
\end{equation}

Indeed, $A\preceq  B$ by definition means that $C(A\cup B)\subseteq  B$, from where
$\widehat{C}(B)\subseteq  \widehat{C}(C(A\cup B))\stackrel{(\ref{inv})}{=}\widehat{C}(A\cup
B)\subseteq \widehat{C}(A)$ (since $A \subseteq  A\cup B$).\medskip

Together with (\ref{app}), this gives the equivalence of the following statements: $A\preceq  B$ and $\widehat{C}(A\cup B)=\widehat{C}(B)$.\medskip

\textbf{Lemma 1.} $A\preceq  B$\emph{ if and only if} $A\cap\widehat{C}(B)\subseteq   B$.\medskip

Proof. $A\preceq  B$ means that $C(A\cup B)\subseteq  B$. In other words, $(A\cup
B)\cap \widehat{C} (A\cup B)\subseteq  B$. In particular, $A\cap\widehat{C}(A\cup B)\subseteq  B$.
It remains to remember that $\widehat{C}(A\cup B)=\widehat{C}(B)$.

Conversely, let $A\cap\widehat{C}(B)\subseteq  B$. Then $C(A\cup B)=(A\cup B)\cap \widehat{C}(A\cup
B)\subseteq  (A\cup B)\cap \widehat{C}( B)=(A\cap \widehat{C} (B))\cup (B\cup \widehat{C}(B))
\subseteq  B\cup
B=B$.  Note that at this point we used the distribution law. $\Box$\medskip

This statement offers a slightly different interpretation of the relation $\preceq B$: any element of $A$ that is desirable in the $B$ state, is already within the set $B$.\medskip

In particular, for any family $(A_i,\i\in I)$ of subsets, we have the inclusion
\begin{equation}\label{int}
D(\cup_i A_i)\subseteq  \cap_i D(A_I).
\end{equation}

It is quite simple to check that if $D$ is a desirability operator (that is, it has two
using the above properties), then the choice function $A\mapsto A\cap D(A)$ is Plott one.

Consequently, these two concepts are essentially equivalent, allowing us to work with either. In particular, the stability of a contract system $S$ can be restated using desirability operators. Below, we will illustrate this for a problem involving two agents.

\section{Stability with two agents}

Let's return to the stability problem in a situation with two agents, Firm and Worker, with a set $X$ of potential contracts  and Plott choice functions $F$ and $W$ on  $X$ correspondingly. In these terms, the stability of a system $S$ can be written as the equality
$$
S=\widehat{F}(S)\cap\widehat{W}(S).
$$
Indeed, the inclusion $S\subseteq \widehat{F}(S)\cap\widehat{W}(S)$ is equivalent to the equalities
$S=F(S)$ and $S=W(S)$ (see (\ref{ch})), which gives a). The reverse inclusion says that if some contract is desirable for Firm and for Worker (in the state $S$), then it has already been concluded, that is, it is in $S$.

The set of stable contract systems is denoted as {\bf  S}. As will become clear, this set is non-empty. Through the use of desirability operators, we can already observe that the set {\bf  S} of stable systems is the set $Fix(\widehat{F}\cap \widehat{W})$ of fixed points of mapping
$\widehat{F}\cap\widehat{W}$ the lattices $2^X$ into itself. This hints at the application of Knaster-Tarski theorem. However, this theorem cannot be applied directly because the mapping $\widehat{F}\cap \widehat{W}$ is antitone,  not monotone. The crucial insight here is the following observation.\medskip

\textbf{Proposition 1.} \emph{If $S$ is stable, then $\widehat W (\widehat F(S))=\widehat W(S)$ and
$S=F(\widehat W(S))$. Symmetrically, $\widehat F (\widehat W(S))=\widehat F(S)$ and
$S=W(\widehat F (S))$.}\medskip

Proof. $S\subseteq \widehat F(S)$, from where (by antitonicity $\widehat W$) $\widehat W
(\widehat F (S))\subseteq  \widehat W (S)$. From here
$$
                           \widehat F (S)\cap \widehat W (\widehat F (S))\subseteq
      \widehat F (S)\cap \widehat W (S)=S.
                                                             $$
Again, the antitonicy of $\widehat W$ gives $\widehat W(S)\subseteq  \widehat W (\widehat F (S)\cap
\widehat W (\widehat F (S))=\widehat W (\widehat F(S))$, which together with $\widehat W (\widehat F (S))\subseteq \widehat W(S)$ gives the equality $\widehat W (\widehat F(S))=\widehat W (S)$. Finally, $W(\widehat F (S))=\widehat F( S)\cap \widehat W (\widehat F( S))=\widehat F( S)\cap \widehat W
(S)=S$.  $\Box$\medskip

This motivates the following\medskip

\textbf{Definition.} A set $A$ is called $F$-\emph{stable} if $A=\widehat F(\widehat W
(A))$. In other words, if $A$ is the fixed point of the operator $\widehat F \circ\widehat W$.
Symmetrically, the set $B$ is called $W$-\emph{stable} if $B=\widehat W(\widehat F
(B))$.\medskip

Obviously, if $A$ is $F$--stable, then $\widehat W(A)$ is $W$-stable, and vice versa. For
a stable $S$ the set $ \widehat F(S)$ is $F$-stable, and $\widehat W(S)$ is $W$--stable. Inversely,\medskip

\textbf{Proposition 2.} \emph{Let $A$ be an $F$-stable set. Define $B=\widehat W(A)$ and
$S=A\cap B$. Then $S$ is stable, $A=\widehat F(S)$ and $B=\widehat W (S)$.}\medskip

It is enough to check the last two equalities.
$$
\widehat W (S)=\widehat W (A\cap B)=\widehat W (A\cap \widehat W (A))=\widehat W (A)=B.
$$
The second equality arises similarly, given that $B$ is a $W$-stable set and $\widehat F
(B)=\widehat F
(\widehat W (A))=A$. $\Box$\medskip

To summarize. Let {\bf  FS} be the set of $F$-stable systems, and {\bf  WS} be the set of $W$-stable systems. We have three reciprocal bijections:
$$
{\bf  S} \stackrel{\widehat F}{\longrightarrow} {\bf  FS} \stackrel{W}{\longrightarrow} {\bf       S},
                   $$
$$
{\bf  S} \stackrel{\widehat W}{\longrightarrow} {\bf  WS} \stackrel{F}{\longrightarrow}  {\bf
      S},
$$
$$
{\bf  FS} \stackrel{\widehat W}{\longrightarrow} {\bf  WS} \stackrel{\widehat
      F}{\longrightarrow} {\bf  FS},
 $$
that can be assembled into the "commutative" diagram
$$
\begin{array}{ccc}
          \textbf{FS} & \rightleftarrows & \textbf{WS} \\
          \uparrow\downarrow &  & \uparrow
\downarrow \\
          \textbf{S} & =\!= & \textbf{S}
        \end{array}
$$
In principle, one can work with any of these sets. For instance, with the set of fixed points  {\bf  FS} of the operator $\widehat{W} \circ \widehat{F}$ on the lattice $2^X$. Since this operator is monotonic, the Knaster-Tarski theorem guarantees the existence of fixed points, ensuring that the sets (\text{FS}) and (\text{S}) are non-empty.

Note that the proof of non-emptiness of {\bf  FS} uses only the antitonicity of the operators $\widehat F$ and $\widehat W$. And to compare {\bf  FS} and {\bf  S}, we also need the "Loeb property"\ (\ref{inv})\ for the operators $\widehat F$ and $\widehat W$.

That's all about existence! According to Tarski, there exists such a set $A$ that $A=\widehat{F}(\widehat{W}(A))$. And then $S:=W(A)=A\cap\widehat{W}(A)$ is a stable system
of contracts. While this may suffice, it's worth briefly touching on related questions:

1) about the order (lattice) structure,

2) on the comparison with Fleiner's proof,

3) about a more constructive proof,

4) about comparative statics,

5) about the case of multiple agents,

6) about a generalization to graduated contracts.

\section{Order structure}

According to Tarski, the set {\bf  FS} of fixed points of the operator $\widehat F\circ\widehat W $
is a lattice subposet of the poset $(2^X,\subseteq  ) $ (although not a sublattice of the lattice $(2^X, \subseteq  $). On the other hand, stable systems (like any other) can be compared from  the positions of Firm (using the Blair relation $\preceq _F$) or from the position of Worker (using $\preceq _W$).

As we will see, these structures actually coincide. More precisely, the mapping $\widehat F :{\bf  S} \to {\bf  FS}$ is an isomorphism of the posets  $({\bf  S},\preceq  _F)$ and $({\bf  F}{\bf  S},\supseteq )$. Symmetrically, $\widehat W :{\bf  S} \to {\bf  W}{\bf  S}$ is an isomorphism of posets $({\bf  S},\preceq  _W)$ and $({\bf  W}{\bf  S}, \supseteq )$.

Together with Tarski, this gives that the poset $({\bf  S},\preceq _F)$ (as well as $({\bf  S},\preceq  _W)$) is a complete lattice. And since the map $\widehat F$ and $\widehat W$ is antitonic, the orders $\preceq  _F$ and $ \preceq  _W$ on the set {\bf  S} are opposite.  In the literature, this is called the \emph{polarization of interests theorem}.

All this can be seen from the following lemma.\medskip

\textbf{Lemma 2.} \emph{Let $S$ and $T$ be stable systems of contracts. The following statements are equivalent:}

        1) $S\preceq  _F T$;

2) $\widehat F (T)\subseteq  \widehat F (S)$;

3) $\widehat W (S)\subseteq  \widehat W (T)$;

4) $T\preceq  _W S$.\medskip

Proof. $1) \Rightarrow  2)$ due to generalized antitonicity (\ref{anti}).

$2) \Rightarrow  4)$. Due to Lemma 1, we need to show the inclusion $T\cap \widehat{W}(S)\subseteq  S$. Since $T\subseteq \widehat{F} (T)\subseteq \widehat{F}(S)$, then $T\cap \widehat{W}(S)\subseteq  \widehat{F}(S)\cap \widehat{W}(S)=S$.

Implications of $4) \Rightarrow 3)$ and $3)\Rightarrow 1)$ are proved the same way. $\Box$\medskip

\section{Comparison with Fleiner}

The idea of using fixed points for stability problems is not new. Back in the works of Feder
and Subrahmanyam in the early 90s, stable matchings were interpreted as fixed points of
suitable mappings.  In relation to the formulation discussed above, this approach and the reduction to Knaster-Tarski theorem was implemented by Fleiner \cite{F}. In general, our scheme of reasoning
is very close to Fleiner's scheme. Therefore, it is worth dwelling on the differences.

The main difference and simplification is related to the use of desirability operators. To obtain
monotonic mappings associated with a Plott choice function $C$, Fleiner turns to the operator
${\overline{C}}$, given by the formula $\overline{C}(A)=A-C(A)$. In other words, these are unselected elements of $A$. This operator is monotone (this is a reformulation of substitutability condition), and it can be used for applications of Tarski's fixed point theorem.

More precisely, Fleiner introduces a notion of a stable $FW$-pair as a pair of sets $(A,B)$ such that $A\cup B=X$ and $FA=A\cap B=WB$. It is quite easy to verify that in this case $A\cap B$ is stable. On the other hand, Tarski's theorem guarantees the existence of stable pairs.

We also obtain stable $S$ through the intersections of the sets $\widehat{F}(S)$ and $\widehat{W}(S)$. These sets differ from Fleiner stable pairs. Formally, their union $\widehat{F} (S)\cup\widehat{W}(S)$ is not equal to $X$. They also differ in meaning. The Fleiner's set $A$ is, roughly speaking, a "budget"\ set for Firm, a set where it can operate freely. And the condition $A\cup B=X$ guarantees non-blockability. Indeed, a blocking element $x$ lies in $A$ or $B$. If in $A$, then since $F(A)=S$, $F(A\cup x)=S$ by virtue of
consistency of $F$, so $x$ is not desirable for Firm in the state $S$.

In our case, $\widehat{F}(S)$ is, on the contrary, a set of desirable contracts for Firm, and the choice of Firm  from this set is not $S$. The choice of Worker $W(\widehat{F}(S))$ is equal to $S$.

Another important difference is that there can be many $FW$ pairs giving the same $S$. At that
while $\widehat{F}(S)$ and $\widehat{W}(S)$ are uniquely and canonically defined by $S$. And one
 half of this pair defines the other. Because of this, we can work with one half and get a bijection (consistent with the orders) of the set {\bf  S} of stable systems and the set {\bf  FS} of
$F$-stable systems. The last one is the set of fixed points of a monotonous mapping
$\widehat{F}\circ\widehat{W}$, which immediately gives the non-emptiness and latticeness of the posets{\bf  FS} and {\bf  S}.

\section{More constructive!}

The above proof looks unconstructive due to the reference to Tarski's theorem, which
looks like a pure statement of existence. However, one has to clearly understood that in the basis of Tarski theorem lies a simple idea is successive approximation, that is, an almost explicit construction. All you need to do is to find a set $A\subseteq  X$ for which $A\subseteq \widehat{F}(\widehat{W}(A))$. Then iterations of the set $A$ using the mapping $\widehat{F}\circ \widehat{W}$ grow monotonously and converge to a fixed point. All looks especially simple in the case when the set $X$ is finite (in the general case, it needs to involve transfinite iterations). In the finite case, the sequence

$$
A\subseteq  (\widehat{F}\circ \widehat{W} )(A) \subseteq  (\widehat{F}\circ \widehat{W})^2 (A))
\subseteq  ...
$$
stabilizes after finite steps and gives ay $F$-stable set $A_\infty$. The corresponding stable set $W(A_\infty)$ is denoted as $\Sigma(A)$. Obviously, $A\preceq _W\Sigma(A)$. Less obviously, that $A$ is the worst stable system for Worker, which is no worse than $A$.\medskip

\textbf{Proposition 3.} \emph{If $T$ is a stable system and $A\preceq _W T$, then $\Sigma(A)
\preceq  _WT$ (and $T\preceq _F\Sigma(T)$).} \medskip

Proof. Since $A\preceq _W T$ then, by virtue of generalized antitonicity (\ref{anti}),
$\widehat{W}(T)\subseteq  \widehat{W}(A)$ and then $\widehat{F}(\widehat{W}(A))\subseteq
\widehat{F} (\widehat{W}(T))=\widehat{F}(T)$. And so on, what gives $A_\infty\subseteq
\widehat{F}(T)$. But $A_\infty =\widehat{F}(\Sigma(A))$, so by Lemma 2 we get
$\Sigma(A)\preceq _W T$. $\Box$\medskip

In particular, if we take $A=\emptyset$, we get as $\Sigma(\emptyset)$ the worst for Worker (and the best for Firm) stable contract system. Let's take a closer look at how this  process works in the case of empty $A$. $\widehat{W}(\emptyset)$ is all $X$, and $\widehat{F}(X)$ is the set of the best contracts for Firm. So in the first step, we receive Firm's proposals: Firm offers to Worker its best contracts. In the second step, Worker selects from these offers the best ones for Worker and rejects "bad"\ offers. From the resulting set, Firm chooses the best ones and offers them to Worker. Etc. In other words, this is actually the Gale-Shapley algorithm \cite{GS}. And the process proposed above differs only in that it can begin not with an empty set, but with any set $A$ satisfying
 the inclusion $A\subseteq \widehat{F}(\widehat{W}(A))$.

The same method allows to build more constructively the supremum $S\vee_W T$ of two stable
$S$ and $T$ with respect to the Worker's order $\preceq _W$. Take $A=S\cup T$. For reasons of monotony, we have $A\subseteq \widehat{F}(\widehat{W}(A))$. Therefore, $\Sigma(A)$ is a minimal stable  system which is $\preceq _W A$, that is, $\preceq  _W$ at the same time $S$ and $T$. That is $\Sigma(A)$ is the supremum of $S$ and $T$.

Another application of this method relates to comparative statics.

\section{Comparative statics}

So far, we have been dealing with a fixed problem $(X;F,W)$. Now let's assume that we have
a "close"\ problem $(X;F',W')$ (again with Plott choice functions $F'$, $W'$ on the same set
of contracts $X$), $F\subseteq  F'$ and (for simplicity) $W=W'$. In other words, in a new problem Firm becomes less demanding (it chooses larger sets), whereas Worker retains its preferences. Intuitively, this should lead to outcomes that are more favorable for Worker. Below we
formulate this prediction more precisely in the style of (Danilov and Koshevoy, 2025).

Let $S$ be a stable system of contracts in the original problem, and $A=\widehat{F}(S)$. This set
$A$ is \textbf{FS}-stable in the original problem, but, generally speaking, not in the modified one with  the new Plott choice function $F'$. Because $\widehat{F'}\supseteq\widehat{F}$. However, for the  same reason, there is the inclusion
$$
A=\widehat{F}(\widehat{W}(A))\subseteq  \widehat{F'}(\widehat{W}(A)).
$$

And we can use this set as the starting point of a natural process (described above, but for the new problem), resulting in a set $S'=W(A_\infty)=\Sigma'(A)$ which is stable in the new problem. So we have a natural way to transform stable systems from the original problem to the stable systems of the new problem. That is, we get the mapping ${\bf  S}\to {\bf S'}$.

As already mentioned, $A\preceq _W S'$. Also, $S\approx_W A$. Therefore, $S\preceq _W S'$, that is, the new stable system $S'$ (obtained in a "natural"\ way from the old stable system $S$), is no worse than $S$ for Worker. And most likely even better. Which, as mentioned above, is to be
expected.

\section{The case of several agents}

So far, we have been working with two agents. Now let's briefly talk about the general case. Consideration of an arbitrary (multi)graph occurs in two stages. In the first stage, we will discuss bipartite graphs, and in the second, general graphs.

\emph{A bipartite case.} Here it is assumed that agents are divided into two groups, conditionally into firms ($f$) and workers ($w$). And contracts are only possible between agents from different
groups. Such a problem can easily be reduced to the case of two agents by aggregating all firms into one Firm and of the workers into one Worker. The Firm's choice selection function has the form $F=\cup _f C_f$ and is set on the set $X=\cup _f X(f,w)=\cup _{f,w} X(f,w)$. Similarly, the aggregated Worker has  choice function $W=\cup _w C_w$ on the same set of $X$.

It is easily verified that $F$ and $W$ are Plott functions of choice, and the set of contracts
$S\subseteq  X$ is stable in an aggregated problem if and only if it is stable in the initial (disaggregated) problem. In particular, the set of stable systems has (two  opposite) lattice structures defined by the intersections of the Blair orders of firms and workers,
respectively.

In the disaggregated formulation, it is possible to concretize the previous discussion of comparative statics. Let's imagine that a new firm $f_0$ appears. Then the aggregated Firm begins to choose  more, and this has a positive effect on the position of workers. More precisely, so that it does not change there are many global contracts $X$, it is more convenient to assume that this "new" firm $f_0$ was present initially, but it didn't want to sign any contracts. That is, its initial choice function $C_{f_0}$ on the set $X(f_0)$ was empty. And in a new problem, it becomes non-
trivial. As was discussed in the previous section, the old stable system is no longer stable because workers have new employment opportunities in this 'new' firm $f_0$. Some of them sign contracts with $f_0$; this usually leads to them abandoning some of the old contracts with other firms. These firms are trying to fill vacant positions and a chain of contract renegotiations begins, leading to a new stable system. For more information, see \cite{Y}.\medskip

\emph{The case of a general graph.} The general case is reduced to a bipartite one by splitting each agent into two halves, a firm and a worker. The resulting "doubled" problem has an obvious symmetry
(permutation of the halves of each agent), which leads to the fact that the set $\widetilde {\bf s}$ of stable solutions of the modified problem have an involution $\tau $ (consistent with order
structures on $\widetilde{\bf  S}$ in that it changes the order to the opposite one). And the set
{\bf  S} of stable solutions to the initial problem are the set of fixed elements of this involution
$\tau$ to $\widetilde {\bf  S}$. This is described in more detail in \cite{Dan-25}.

\section{Generalization to graduated contracts}

The approach to stability theory presented above can be extended to a more general scenario where contracts can be concluded with intermediate intensity, ranging between complete renunciation and full acceptance \cite{AG, Dan-21}. This means contracts can be graded. For instance, a deposit agreement, besides interest and term, specifies a variable amount of money that can range, say, from 0 to 10 million rubles. Similarly, an employment contract might include variable working hours.

Formally, this implies that for every contract $ e \in X $, there exists a specific scale $ L(e) $. This scale is a linearly ordered set representing the "levels of performance"\ for contract $e$. It's reasonable to assume this scale forms a complete chain, with the minimum element signifying refusal to conclude the contract. For simplicity, we assume each scale is a finite chain.

A contract system can now be viewed as a family of elements $x_e \in L_e$ for every $e \in X$, that is as an element of the Cartesian product $L = \prod_e L_e$. This set, equipped with the coordinate-wise order $\leq$, is a complete distributive lattice. In simpler terms, a contract system is akin to a function defined over the set $X$. Stability is defined as a property where each participant can reduce their level of involvement in the contract unilaterally, but can only increase it through mutual agreement.

A choice function in this setting is a mapping $C:L\to L$ such that $C(x)\le  x$ for any
$x\in L$. It is assumed that choice functions of all participants satisfy the conditions of consistency and substitutability, which are formulated as before with the replacement of the icons $\subseteq , \cap, \cup$ by $\le , \wedge, \vee$. As above, the Blair relation $
\preceq =\preceq  _C$ on the lattice $L$ is determined by
$$
x\preceq  y, \text{ if } C(x\vee y)\le  y.
$$
As above, the desirability operator $\widehat{C} :L\to L$ is introduced. More precisely, an element $x\in L$ it is called \emph {desirable} in the state $a\in L$ if $x\le  C(a\vee x)$. The supremum of all desirable elements (in a state $a$) is denoted as $\widehat{C}(a)$.

The properties and connections of these concepts are basically the same as in the set-theoretic case, and the proofs follow the same path. Therefore, we will limit ourselves to just the wording.

1) $C (\vee_i  x_i)=C(\vee_i C(x_i)$ for any family of $x_i$.

2) the operator $\widehat{C}$ is antitone. Moreover, if $a\preceq  b$, then $\widehat{C}(b)\le
\widehat{C}(a)$.

3) $C(a)=a\wedge\widehat{C}(a)$ for any $a\in L$. In particular, $C(a)\le \widehat{C}(a)$.

Let's explain this in more detail. First of all, note that the element $C(a)$ is desirable in the state $a$, so that $C(a)\le \widehat{C}(a)$. Therefore, $C(a)\le  a\wedge\widehat{C}(a)$, and it remains to establish the opposite inequality $C(a)\ge a\wedge\widehat{C}(a)$. To do this, we need to show that for any desirable element $z$ in the state $a$, the inequality $C(a)\ge a\wedge z$ is satisfied. And  this follows from an application of the substitutability property to the inequality $a\le  a\vee z$.

Thus, the operator $\widehat{C}$ has the formal properties of a desirability operator from Section 2.

4) If $z$ is desirable in the state of $a$, and $y\le  z$, then $y$ is also desirable.

Indeed, $a\vee y\le  a\vee z$, so from the substitutability we have $(a\vee y)\wedge C(a\vee z)\le  C(a\vee y)$. But $y\le  a\vee y$, and $z\le  C(a\vee z)$, so $y=y\wedge z\le  C(a\vee y)$.\medskip

Let us now turn to the stability problem with two agents, Firm and Worker. Their choice functions on the lattice $L$ of graded contracts are denoted by $F$ and $W$, respectively. A contract system $s\in L$ is called \emph{stable} if (as before) $s= F(s)=W(s)$, and if an element $z\in L$ is desirable in the state $s$ for Firm and Worker, then $z\le  s$.\medskip

\textbf{Lemma 3.} \emph{An element $s$ is stable if and only if $s=\widehat{F}(s)\wedge \widehat{W}(s)$.} \medskip

Proof. Let $s$ be stable. Then $s=F(s)\le \widehat{F}(s)$ and similarly for $W$, whence (see 3)) $s\le  \widehat{F}(s)\wedge\widehat{W}(s)$. Now let's assume that $s< \widehat{F}(s)\wedge \widehat{W}(s)$. This means that for a certain type of contract $e\in X$, a strict inequality holds
$s(e)< \widehat{F}(s)(e)$ and similarly for $W$. This means that there is a $z$ desired for
Firm (in the state of $s$) such that $s(e)<z(e)$. Similarly, there is a $y$ desirable for Worker
and such that $s(e)<y(e)$. Moving on to $z\wedge y$ and using the property 4), we can assume that
$y=z$. So, this $z$ is desirable for both Firm and Worker, and at the same time, $z\nleq s$, which contradicts  to stability.

Conversely, let $s=\widehat{F}(s)\wedge\widehat{W}(s)$. Then $s\le \widehat{F}(s)$ and, by the property 3), $s=F (s)$. Similarly for $W$. It remains to check the nonblockability. Let $z$ be desirable for Firm and for Worker in the state $s$. Then $z\le \widehat{F}(s)$ and $s\le \widehat{W}(s)$, from where $z\le s$. $\Box$\medskip

Now it's clear that all previous statements about stability still apply in this "lattice" framework. This is because the analysis is based on properties of the operators $ F, \widehat{F}, W$  and  $\widehat{W}$/. These properties remain the same in the more general situation. Consequently, in this broader setting, we retain the fundamental findings: existence, polarization, and latticeness.

\end{document}